\documentclass[12pt]{article}
\def\be{\begin{equation}}
\def\ee{\end{equation}}

\begin{document}

\title{Noncommutative probability in classical systems}

\author{Andrei Khrennikov
\\International Center for Mathematical
Modeling \\ in Physics and Cognitive Sciences,\\
University of V\"axj\"o, S-35195, Sweden\\
S.V.Kozyrev\\
Institute of Chemical Physics, Russian Academy of Science, Moscow}

\maketitle

\begin{abstract}
Two examples of the situation when the classical observables
should be described by a noncommutative probability space are
investigated. Possible experimental approach to find quantum-like
correlations for classical disordered systems is discussed. The
interpretation of noncommutative probability in experiments with
classical systems
as a result of context (complex of experimental physical conditions) 
dependence of probability is considered.
\end{abstract}

\section{Introduction}

It is widely believed, that classical systems should be described
by commutative, or Kolmogorovian, probability space. In the
present paper we investigate the following question: is it
possible to observe correlation of noncommutative observables in
purely classical situation? This would mean, that the classical
system will be described by noncommutative (or quantum)
probability space.

The related subjects were investigated in Bohmian mechanics
\cite{Bohmian}, \cite{Bohmian1} and in approach by t'Hooft \cite{tHooft}, \cite{tHooft1}, 
where the
properties of quantum system were discussed as a result of some
underlying classical dynamics in the space of hidden classical
parameters. In papers by Accardi and Regoli, see \cite{AR},
violation of the Bell inequality in classical system was
discussed.

Another related subject was discussed in papers \cite{KhVol},
\cite{KhVol1}, where numerical simulation of discretized classical
mechanics was performed, and quantum like interference fringes
were obtained.

In  papers \cite{KHR1}, \cite{KHR2} of one of the authors there was developed
a contextual probabilistic approach to the statistical theory 
of measurements over quantum as well as classical physical systems.
It was demonstrated that by taking into account dependence of probabilities on
complexes of experimental physical conditions, {\it physical contexts,}
we can derive quantum interference for probabilities of alternatives.
Such a contextual derivation is not directly related to special 
quantum (e.g. superposition) features of physical systems. Those contextual 
models \cite{KHR1}, \cite{KHR2} stimulated the present search for noncommutative structures in classical statistical
physical models, e.g. disordered systems.

In the present paper we discuss two examples of {\it classical
statistical mechanical systems} where we obtain the correlation
functions in {\it noncommutative probability space.}

Our approach, in principle, may be compared with experiments in
the following standard way. The traditional way to distinguish
between classical and quantum system is the Bell inequality,
satisfied by classical correlation functions. Thus if we would
find violation of the Bell inequality, we will prove that the
system is described by noncommutative probability space. However,
our present theoretical considerations are still far from such
experimental study.

The structure of the present paper is as follows.

In Section 2 we investigate the example of arising of
noncommutative probability for classical observables as a result
of time averaging.

In Section 3 we consider the correlation functions on
noncommutative probability space for classical disordered system.
The noncommutativity there will be a result of ensemble averaging.
Also in this Section we discuss the considered examples from the
point of view of the context dependent interpretation of
noncommutative probability.

\section{Noncommutative probability and time averaging}

In the present section we discuss the following problem. Consider
the dynamics of quantum system, described by some Hamiltonian
$H_0$ and the algebra of observables ${\cal A}$. Let this
(noncommutative) algebra of observables ${\cal A}$ contains some
(commutative) classical subalgebra ${\cal C}$. This classical
subalgebra is not conserved by time evolution, but for $X$,
$Y\in{\cal C}$ the time evolutions $X(t)=e^{itH_0}Xe^{-itH_0}$ and
$Y(t)=e^{itH_0}Ye^{-itH_0}$ will commute by definition.

Let us assume that the time evolution, defined by Hamiltonian
$H_0$ is very fast, and in experiment we observe some time
averaged observables. These time averaged observables, in general,
already will not commute, since the classical subalgebra is not
conserved by time evolution. This means that, in principle, we
might expect that these time averaged operators for classical
physical variables $X(t)$, $Y(t)$ from the classical subalgebra
${\cal C}$ will have nonclassical correlations.

The natural example of this kind of behavior is observed in the
quantum stochastic limit approach \cite{book}. In this approach we
consider the quantum system with the Hamiltonian in the form
$$
H=H_0+\lambda H_I
$$
where $H_0$ is called the free Hamiltonian, $H_I$ is called the
interaction Hamiltonian, and $\lambda\in {\bf R}$ is the coupling
constant.

We investigate the dynamics of the system in the new slow time
scale of the stochastic limit, taking the van Hove time rescaling
\cite{vanHove}
$$
t\mapsto t/\lambda^2
$$
and considering the limit $\lambda\to 0$. In this limit
\cite{book} the free evolutions of the suitable collective
operators
$$
A(t,k)=e^{itH_0}A(k)e^{-itH_0}
$$
will become quantum white noises:
$$
\lim_{\lambda\to 0}{1\over\lambda}A\left({t\over
\lambda^2},k\right)=b(t,k)
$$
The convergence is understood in the sense of correlators. The
$\lambda\to 0$ limit describes the time averaging over
infinitesimal intervals of time and allows to investigate the
dynamics on large time scale, where the effects of interaction
with the small coupling constant $\lambda$ are important.

For the details of the procedure see \cite{book}.

The collective operators describe joint excitations of different
degrees of freedom in systems with interaction, and may have the
form of polynomials over creations and annihilation of the field,
or may look like combinations of the field and particles operators
etc.

For example, for nonrelativistic quantum electrodynamics without
the dipole approximation the collective operator is \be\label{QED}
A_j(k)=e^{ikq}a_j(k) \ee where $a_j(k)$ is the annihilation of the
electromagnetic (Bose) field with wave vector $k$ and polarization
$j$, $q=(q_1,q_2,q_3)$ is the position operator of quantum
particle (say electron), $qk=\sum_{i}q_ik_i$.

The nontrivial fact is that, after the $\lambda\to 0$ limit,
depending on the form of the collective operator, the statistics
of the noise $b(t,k)$ depends on the form of the collective
operator and may be nontrivial.

Consider the following examples.

1) We may have the following possibility \be\label{Bose}
[b_i(t,k),b_j^{\dag}(t',k')]=2\pi\delta_{ij}\delta(t-t')\delta(k-k')\delta(\omega(k)-\omega_0)
\ee which corresponds to the quantum electrodynamics in the dipole
approximation, describing the interaction of the electromagnetic
field with two level atom with the level spacing (energy
difference of the levels) equal to $\omega_0$. Here $\omega(k)$ is
the dispersion of quantum field.

In this case the quantum noise will have the Bose statistics, and
different annihilations of the noise will commute
$$
[b_i(t,k),b_j(t',k')]=0
$$

2) The another possibility is the relation \be\label{qB}
b_i(t,k)b_j^{\dag}(t',k')=2\pi\delta_{ij}\delta(t-t')\delta(k-k')\delta(\omega(k)+\varepsilon(p)-\varepsilon(p+k))
\ee which corresponds to the quantum electrodynamics without the
dipole approximation (\ref{QED}). Here $\omega(k)$ and
$\varepsilon(p)$ are dispersion functions of the field and of the
particle correspondingly.

In this case the quantum noise will have the quantum Boltzmann
statistics \cite{book}, \cite{AcLuVo}, \cite{QED}, and different
annihilations of the noise will not commute
$$
b_i(t,k) b_j(t',k')\ne b_j(t',k') b_i(t,k)
$$

The commutation relations of the types (\ref{Bose}), (\ref{qB})
are universal in the stochastic limit approach (a lot of systems
will have similar relations in the stochastic limit $\lambda\to 0
$).

Take two operators $b_i(t,k)$ for the same time and fixed
polarization and consider the combinations \be\label{combo}
X(k)=b_i(t,k)+b_i^{\dag}(t,k),\qquad
X(k')=b_i(t,k')+b_i^{\dag}(t,k') \ee which correspond to the
coordinate operator.

Then for the case of quantum Boltzmann relations(\ref{qB}) we have
$$
X(k)X(k')\ne X(k')X(k)
$$
Of course, we need some regularization of the product of
generalized functions.

Operators $X(k)$ before the stochastic limit belonged to the
classical subalgebra. More precisely, corresponding combinations
$x(k)$ of interacting operators (\ref{QED}) would belong to the
classical subalgebra:
$$
x(k)=A_i(k)+A^{\dag}_i(k)
$$
$$
[x(k),x(k')]=0
$$

We proved, that for the case when after the stochastic limit the
statistics of the field become quantum Boltzmannian, operators
(\ref{combo}) will not commute even if we take them for equal
time. This is not a mystery, since in the stochastic limit we work
with the time averaged observables. But in real experiments we may
observe the result of time averaging. The discussed example shows,
that for quantum electrodynamics beyond the dipole approximation
we may observe quantum correlations for time averaged observables
in the classical subalgebra.

The considered in the present section situation is similar in some
sense to the results of \cite{KhVol}, \cite{KhVol1}. In these
papers the quantum like interference fringes were observed for the
discretization of the classical dynamical system. The
discretization is an analog, in some sense, of the time averaging
procedure. Probably the results of \cite{KhVol}, \cite{KhVol1}
might be possible to embed into the frameworks of the approach
considered in the present section.

We would like to mention that since long time De Muynck, 
see e.g.  \cite{dM},  discuss analogy between thermodynamics and
quantum mechanics. They considered in the EPR--Bohm framework
quantum expectations as a kind of thermodynamic averages. This
should induce violation of Bell's inequality.

\section{Noncommutative probability and disordered systems}

In the present section we discuss the possibility of using of
noncommutative probability to describe (classical) disordered
systems, following \cite{nra}, \cite{nrp}. In these papers the new
procedure, called the noncommutative replica procedure, which is
an analog of the replica procedure of Edwards and Anderson
\cite{EA}, was proposed to describe the statistical mechanics of
quenched disordered systems (for example, spin glasses).

We will not discuss here the standard replica approach, see for
introduction to spin glasses and the replica method \cite{EA},
\cite{SpinGlass}, \cite{BY}.

Consider the disordered system with Hamiltonian $H[\sigma,J]$
which depends on the random parameter $J$ which in the most
interesting cases (for spin glasses for instance) is the large
random $N\times N$ matrix with independent Gaussian matrix
elements $J_{ij}$, considered in the thermodynamic $N\to\infty$
limit.

To describe the system with quenched disorder in \cite{nra},
\cite{nrp} it was to proposed to consider the state described by
the noncommutative replica statistic sum \be\label{ncreplica1}
Z^{(p)}=\int\sum_{\{\sigma\}} \exp\left(-\beta H[\sigma,\Delta
J]\right) \prod_{a=0}^{p-1}\exp\left(-{1\over2}\sum_{i\le j}^N
J^{(a)2}_{ij}\right) \prod_{i\le j}^N dJ^{(a)}_{ij} \ee where
$\Delta$ is the following coproduct operation \be\label{Delta}
\Delta: J_{ij}\mapsto {1\over\sqrt{p}}\sum_{a=0}^{p-1}J_{ij}^{(a)}
\ee which maps the matrix element $J_{ij}$ into the linear
combination of independent replicas $J_{ij}^{(a)}$, enumerated by
the replica index $a$. This operation was called the quenching in
\cite{nrp}.

In the large $N$ limit, by the Wigner theorem, see
\cite{Wig}--\cite{ALV}, the system of $p$ random matrices with
independent variables will give rise to the quantum Boltzmann
algebra with $p$ degrees of freedom with the generators $A_a$,
$A_a^{\dag}$, $a=0,\dots,p-1$ and the relations
$$
A_aA_b^{\dag}=\delta_{ab}
$$
These operators are the limits of the large random matrices
$$
\lim_{N\to\infty}{1\over N}J_{ij}^{(a)}=Q_a=A_a+A_a^{\dag}
$$
where the convergence is understood in the sense of correlators
(as in the central limit theorem).

Then in the thermodynamic limit $N\to\infty$ the noncommutative
replica procedure (\ref{Delta}) will take the form of the
following map of the quantum Boltzmann algebra with one degree of
freedom into quantum Boltzmann algebra with $p$ degrees of
freedom:
$$
\Delta: Q\mapsto {1\over\sqrt{p}}\sum_{a=0}Q_a
$$

Note that different $Q_a$ do not commute. We see, that we again
have obtained noncommutative probability in purely classical
system.

Actually the picture is more complicated, compared to the
discussed above. The correlations of the system in the
noncommutative replica approach will be given by
$$
\lim_{N\to\infty}\langle \left(\Delta J\right)^k\rangle=\langle
\left({1\over\sqrt{p}}\sum_{a=0}Q_a\right)^k\rangle
$$
where the state $\langle\cdot\rangle$ is generated by the
noncommutative replica statistic sum (\ref{ncreplica1}). In
principle, it is not clear, how to extract noncommutativity from
this set of correlators, since different degrees $\left(\Delta
Q\right)^k$ commute.

To distinguish noncommutative and commutative systems we have to
consider the set of correlation functions which will be large
enough. In the present case this set should contain correlations
of different linear combinations of $Q_a$, more general than
$\Delta Q$.

This problem may be discussed in the following way. The quenching
(\ref{Delta}) in principle may be related to particular way of
preparation of the disordered system under consideration. If we
will use different physical preparation of the disordered system,
this may result in the different quenching procedure. The example
of quenching different from (\ref{Delta}) was discussed in
\cite{nrp}. This example has the following form \be\label{onX1}
\Delta': J\mapsto {1\over\sqrt{p}}\sum_{a=0}^{p-1} c_a J_a; \ee
where $c_a$ are real valued coefficients, which should satisfy the
condition
$$
\sum_{a=0}^{p-1} c_a^2 =p
$$
Varying coefficients $c_a$ we will obtain different quenchings.

Then, using different physical preparations of the system, we
measure the correlation functions which will correspond to
different quenchings. After we may (at least in principle) use the
Bell inequality to prove, do we really have noncommutative
probability space which describes the behavior of the disordered
system under investigation.

Actually the most natural example of this setup is the experiments
with spin glasses, where the glass transition with different
external magnetic field was investigated, and non trivial behavior
of magnetization on preparation was observed \cite{SpinGlass},
\cite{BY}.

We propose to analyze these experiments taking into account the
correlations between the systems with different preparations (i.e.
freezed in the presence of different external magnetic fields,
which in our approach should correspond to different quenchings),
and to check the validity of the Bell inequalities.

This would help to check the validity of the noncommutative
replica approach itself, since there is no direct way to introduce
noncommutativity in the standard replica approach.

The discussed here experimental situation could also be described
by using contextual probabilistic approach, see \cite{KHR1}, \cite{KHR2}.
In the contextual framework probabilities (which are interpreted
as conventional ensemble probabilities) depend on physical
contexts --- complexes of experimental physical conditions.
Mathematically this means that we could not use one fixed
Kolmogorov probability space and we should work with a system of
probability spaces depending on physical contexts. In our case
various contexts are defined by choosing various external magnetic
fields. We recall that by using contextual probabilistic approach
it is possible to obtain ''quantum rule'' for interference of
probabilities of alternatives, see \cite{KHR1}, \cite{KHR2}. Such an
interference of probabilities is just another way to describe
noncommutativity of probabilities (induced in the conventional
formalism by using Hilbert space calculus). Therefore we could
expect that, by taking into account contextuality of statistics
for disordered systems freezed in the presence of different
external magnetic fields, it would be possible to find
experimental confirmations of the presence of noncommutative
structure for classical disordered systems (in particular, spin
glasses).

\bigskip

\centerline{\bf Acknowledgements}

The authors would like to thank L.Accardi and I.V.Volovich
L. Ballentine, S. Albeverio, 
S. Gudder,W. De Muynck, J. Summhammer, P. Lahti, 
A. Holevo,  B. Hiley for fruitful (and rather critical) discussions. 

One of the authors (S.K.) has been partly supported
by INTAS YSF 2002--160 F2, CRDF (grant UM1--2421--KV--02), The
Russian Foundation for Basic Research (project 02-01-01084).
The paper was supported by the grant of The Swedish Royal Academy
of Sciences on collaboration with scientists of former Soviet
Union and EU-network on quantum probability and applications.


\begin{thebibliography}{99}

\bibitem{Bohmian} D. Bohm,   {\it Quantum theory, Prentice-Hall.} 
Englewood Cliffs, New-Jersey, 1951.

\bibitem{Bohmian1} D. Bohm   and B. Hiley, {\it The undivided universe:
an ontological interpretation of quantum mechanics.}
Routledge and Kegan Paul, London,  1993.

\bibitem{tHooft} G. `t Hooft, {\it How does God play dice? (Pre)-determinism at the Planck scale.}
Preprint hep-th/0104219, 2001.

\bibitem{tHooft1} G. `t Hooft, {\it Quantum mechanics and determinism.} Preprint hep-th/0105105, 2001.


\bibitem{KHR1} A. Yu. Khrennikov, Ensemble fluctuations and the origin of quantum probabilistic rule,
J. Math. Phys. 43 (2002) 789-802.


\bibitem{KHR2} A . Yu. Khrennikov, Quantum statistics via perturbation effects of preparation procedures,
Il Nuovo Cimento,  B 117,  (2002) 267-281.


\bibitem{AR} L. Accardi,  M. Regoli, Locality and Bells Inequality, in PQ-QP:
Quantum Probability and White Noise Analysis, 13: Foundations
of Probability and Physics, edited by A.Khrennikov, Proc. of the
Conference in V\"axj\"o, Sweden, World Scientific, Singapore, 2001.

\bibitem{KhVol}A. Yu. Khrennikov, Ja. I. Volovich, Interference effect for probability distributions
of determinsitic particles. Proc. Int. Conf. Quantum Theory: Reconsideration
of Foundations. Ser. Math. Modelling in Phys., Engin., and Cogn. Sc., 455-462,
V\"axj\"o Univ. Press,  2002.http://xxx.lanl.gov/abs/quant-ph/0111159

\bibitem{KhVol1} A.Yu.Khrennikov, Ya.I.Volovich, Discrete time leads to
quantum--like interference of deterministic particles,
Proc. Int. Conf. "Quantum Theory: Reconsideration
of Foundations". Ser. Math. Modelling in Phys., Engin., and Cogn. Sc., 441-454,
V\"axj\"o Univ. Press, 2002. http://xxx.lanl.gov/abs/quant-ph/0203009.


\bibitem{book} L.Accardi, Y.G.Lu,  I.V.Volovich,  Quantum Theory and its
Stochastic Limit, Texts and Monographs in
Physics, Springer Verlag,  2002.

\bibitem{vanHove}   L. Van Hove,  Physica 21 (1955) 617.

\bibitem{AcLuVo}
L. Accardi, Y. G. Lu,  I. V. Volovich, Interacting Fock spaces and
Hilbert module extensions of the Heisenberg commutation
relations, Publications of IIAS, Kyoto, 1997.

\bibitem{QED}  L. Accardi, S. V. Kozyrev, I. V. Volovich,
Dynamical $q$-deformation in quantum theory and the stochastic
limit, J.Phys.A 32 (1999) 3485--3495,


\bibitem{nra} S. V. Kozyrev, The noncommutative replica approach,
http://xxx.lanl.gov/abs/cond-mat/0110238.

\bibitem{nrp} S. V. Kozyrev, The noncommutative replica procedure,
http://xxx.lanl.gov/abs/cond-mat/0210196.

\bibitem{EA}  S. F. Edwards, P. W. Anderson, J.Phys. F5 (1975) 965.

\bibitem{SpinGlass}  M. Mezard,  G. Parisi,  M. Virasoro,
Spin-Glass Theory and Beyond, Singapore, World Scientific, 1987.

\bibitem{BY}  K. Binder, A. P. Young, Spin glasses: experimental
facts, theoretical concepts, and open questions, Reviews of modern
physics {\bf 58} N4 801--976.

\bibitem{Wig}  E. P. Wigner, Characteristic vectors of bordered matrices
with infinite dimensions, Ann. of Math. 62 (1955) 548--564.

\bibitem{Voi92}  D. Voiculescu,  K. J. Dykema,  A. Nica, Free random
variables. CRM Monograph Series, 1, American Math. Soc.,
1992.

\bibitem{AV}   I. Ya. Aref'eva,  I. V. Volovich,
The Master Field for QCD and $q$--deformed Quantum
Field Theory, Nuclear Physics B 462 (1996) 600--612.

\bibitem{ALV} L. Accardi, Y. G. Lu, I. V. Volovich, Noncommutative
(Quantum) Probability, Master Fields and Stochastic Bosonisation,
http://xxx.lanl.gov/abs/hep-th/9412241.

\bibitem{dM} W. M. De Muynck, Interpretations  of quantum mechanics,
and interpretations of violation of Bell's inequality, in PQ-QP:
Quantum Probability and White Noise Analysis, 13: Foundations
of Probability and Physics, edited by A.Khrennikov, Proc. of the
Conference in V\"axj\"o, Sweden, World Scientific, Singapore, 2001.

\end{thebibliography}
\end{document}